\documentclass[fleqn,10pt,twocolumn]{AROB-ISBC-SWARM22}

\title{Investigating the impact of free energy based behavior on human \\in human-agent interaction}

\author{Kazuya Horibe${}^{\dagger \ast}$, Yuanxiang Fan${}^{\ast}$, Yutaka Nakamura${}^{}$ and Hiroshi Ishiguro${}^{}$}
\speaker{KH}
\thanks{\ast These}

\affils{${}$Department of Systems Innovation, Osaka University, Osaka, Japan\\
(E-mail: horibe@irl.sys.es.osaka-u.ac.jp)\\
${}^{\ast}$ These authors contributed equally to this work.}

\abstract{Humans communicate non-verbally by sharing physical rhythms, such as nodding and gestures, to involve each other. This sharing of physicality creates a sense of unity and makes humans feel involved with others. In this paper, we developed a new body motion generation system based on the free-energy principle (FEP), which not only responds passively but also prompts human actions. The proposed system consists of two modules, the sampling module, and the motion selection module. We conducted a subjective experiment to evaluate the "feeling of interacting with the agent" of the FEP based behavior. The results suggested that FEP based behaviors show more "feeling of interacting with the agent". Furthermore, we confirmed that the agent's gestures elicited subject gestures. This result not only reinforces the impression of feeling interaction but could also realization of agents that encourage people to change their behavior. }

\usepackage{subfigure}
\usepackage{caption}
\usepackage{comment}
\usepackage{amsmath}
\begin{document}

\maketitle


\section{Introduction}
People interact with each other by drawing in each other's physical rhythms, such as non-verbal gestures\cite{efron1941gesture,mcneill1985so,mcneill2011hand}. The synchronization of physical rhythms through attraction creates a sense of unity in people and plays an important role in facilitating dialogue in terms of interpersonal impressions and conversation satisfaction\cite{kelly1999offering}. The sense of unity during a dialogue has been quantified by measuring the feature values of the movements of two people during a dialogue. It has been shown that the number of dimensions of the feature values of the movements of two people decreases during a dialogue, suggesting that it is easier to predict each other's movements during a dialogue\cite{nishimura2020human}.

Agents that interact with people daily have been studied extensively for purposes such as guiding people at the reception, serving customers, and recommending products in stores, and household pets\cite{castellano2013multimodal,bickmore2011relational,dermouche2019generative}. Since most agents use the other person's state as input for deciding their next action, they do not recognize their movements and the other person's movements as a set, and it is difficult to obtain the sense of unity that can be achieved through the mutual directional attraction between people through agent-human interaction.

In this study, we designed an agent that generates the next gesture using both its own and the other person's movements. To generate the gesture, we used Friston's free energy principle\cite{friston2015active}. The agent trained its generative model using the gestures of two people who were facing each other and interacting beforehand and generated gestures that took into account both its own and the other person's movements. As a result, the subjects' impressions of the life-likeness and human-like nature of the agent were improved. Not only that, we showed that the subjects paid attention to the agent's gestures and were drawn into the interaction. This proposed model is expected to open the door to a new type of human-agent interaction that actively engages people and encourages them to change their behavior.

\section{Free energy principle based action selection for input-output Hidden Markov Model}

\begin{figure}[htpb!]
    \centering
    \includegraphics[width=1\linewidth]{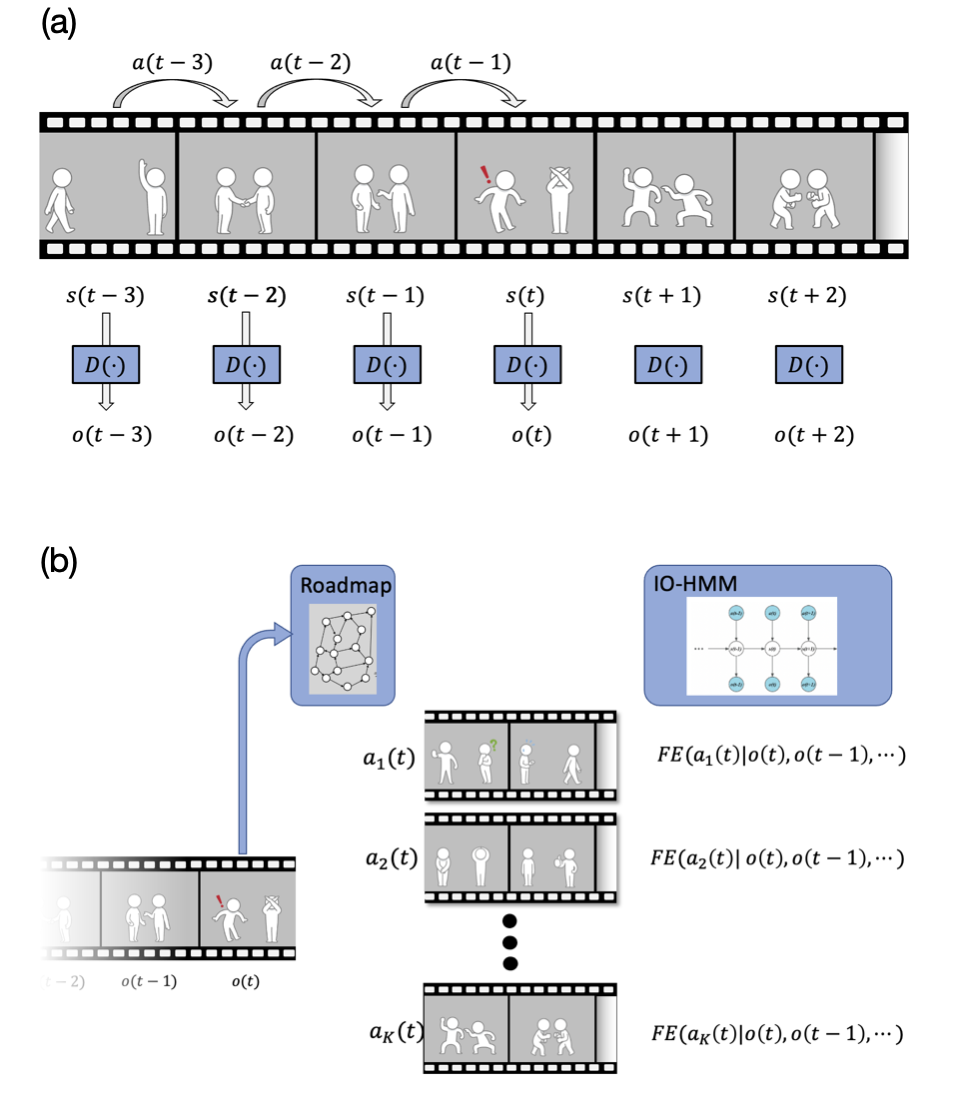}
    \caption{Schematic of the gesture generation model. (a) Action selection module, where D is a ``identifier'' function that returns a decision from the current state s(t) as to whether or not it is interacting with the person in front. (b) Action alternatives are randomly sampled from a roadmap of action sequences. The free energy of each sampled series is calculated and the smallest action series is selected.}
    \label{fig:Sysyem}
\end{figure}

\begin{table*}[h]
  \centering
  \begin{tabular}{ll}
    \hline
    Symbol  & Definition \\
    \hline \hline
    Actions $A(t)$  & $A(t)$ = [a(1), a(2), ..., a(t)], the input sequence from time step 1 to t\\
    Latent state $S(t)$  &  $S(t)$ = [s(1), s(2), ..., s(t)], the latent state sequence from time step 1 to t. \\
    Outputs $O(t)$  & $O(t)$ = [o(1), o(2), ..., o(t)], the output sequence from time step 1 to t \\
    Likelihood $L(\theta, O(t), A(t))$  & The likelihood of IO-HMM giving the output sequence $O(t)$ and the input sequence $A(t)$\\
    IO-HMM parameters $\theta$ & $\theta$ = ($\theta_{IN}$, $\theta_{TR}$, $\theta_{EM}$), which are the parameters of initial, transition, and emission model \\
    \hline
  \end{tabular}
  \caption{Symbol definition of IO-HMM}
  \label{table:data_type}
\end{table*}

In this chapter, we describe the gesture generation model of the proposed agent (Fig.~\ref{fig:Sysyem}). The gesture generation model consists of two components. (1) Sampling module: a probabilistic roadmap to generate gesture sequences (Fig.~\ref{fig:Sysyem}(a)). (2) Action selection module: a time evolution model that calculates the free energy of the series obtained from the gesture roadmap and selects the gesture with the minimum free energy modeled by Partially Observable Markov Decision Processes (POMDP) (Fig.~\ref{fig:Sysyem}(b)). The POMDP modeled the time transition with IO-HMM (Table.~\ref{table:data_type}) and used free energy as a strategy for action selection.






\subsection{Action selection module: Free Eenegy Principle based action selection for IO-HMM}
To inclement POMDP model for generating motion, we proposed to combine the free energy principle (FEP) and Input-output Hidden Markov Model (IO-HMM). For a set of possible actions, the free energy $F$ is calculated for each action, and the policy is to select the action with minimal free energy. The architecture of IO-HMM is shown in Fig. \ref{fig:Sysyem}(b) right side. The blue nodes represent the observed variables, while the white nodes represent the latent variables. The top row contains the input variables a(t); The middle row contains the latent variables s(t); The bottom row contains the output variables o(t). The output of IO-HMM is the free energy, i.e., the entropy of the predicted state at the next time, given the belief state calculated from the observed series so far and an action.

The dynamics is solved by maximizing parameter likelihood L($\theta$,O(T), A(T)) after input data sequence A(T) and output data sequence O(T) has been given. The likelihood can be calculated by

\begin{equation}
\begin{split}
        L(\theta,O(T),A(T)) &= \sum_s Pr(s(1)|a(1);\theta_{IN}) \\ &\prod^T_{t=2}Pr(s(t-1),a(t);\theta_{TR})\\ &\prod^T_{t=2}Pr(o(t)|s(t);\theta_{EM}),
\end{split}
\end{equation}
where $\theta_{IN}$, $\theta_{TR}$ and $\theta_{EM}$ denotes the parameters of initial model Pr($s(1)|a(1)$; $\theta_{IN}$), transition model Pr($s(t)|s(t-1)$, a(t); $\theta_{TR})$ and emission model Pr($o(t)|s(t)$; $\theta_{EM}$) respectively.

The initial probability model is defined as
\begin{equation}
    Pr(s(1)=s_i|a(1);\theta_{IN})=\frac{e^{\theta^i_{IN}a(1)}}{\sum_k e^{\theta^k_{IN}a(1)}}.
\end{equation}

The parameter $\theta_{IN}$ for initial model is a matrix with the ith row $\theta_{IN}^i$ being the coefficients for the initial state being in state $s_i$ and is modeled by using multinomial logistic regression. For a is defined as a time sequence data, $\theta_{IN}^i a(1)$ represented the inner product of features transformed from a(1) and vector $\theta_{IN}^i$.

The transition probability model is defined as
\begin{equation}
\begin{split}
        &Pr(x(t)=x_j|x(t-1)=x_i,u(t);\theta_{TR})= \\ &{e^{\theta^{ij}_{TR}a(t)}}{\sum_k e^{\theta^{ik}_{TR}u(t)}}.
\end{split}
\end{equation}
The parameter $\theta_{TR}$ for transition model is a set of matrices with the jth row of the ith matrix $\theta^{i j}_{TR}$ being the coefficients for the next start being in state $s_j$ given the current state being in $s_i$ and modeled using multinomial logistic regression. For a is defined as a time sequence data,$\theta^{i j}_{TR} a(t)$ represented the inner product of features transformed from a(t) and vector $\theta^{i j}_{TR}$.

The emission probability model is defined as
\begin{equation}
    Pr(o(t)=1|s(t)=s_i;\theta_{EM})=\frac{1}{1+e^{-\theta^i_{EM}}}.
\end{equation}
The parameter $\theta_{EM}$ for emission model is a set of array where $\theta^i_{EM}$  denote the coefficients when the hidden state is $s_i$.

\subsection{Sampling module:Probabilistic roadmap for motion sampling}
To obtain a choice of action sequences, we used a probabilistic roadmap for motion sampling (Fig. \ref{fig:Sysyem}(b) left side). one node represents the posture of the upper body, and the edges represent the probability of transitioning from one posture to another. A single node represents an upper body posture, and an edge represents the probability of transition from one posture to another. The choices of action sequences are randomly sampled action sequences of a fixed time step.

\section{Data collection and \\ module training}
\subsection{Interaction motion data collection}
We recorded by an omnidirectional camera(Insta 360 Air, Shenzhen Arashi Vision, China). The participants are asked to watch an interesting video then talk about it with the experimenters one by one. An omnidirectional camera between them is used to record video data of participants and experimenters. The video data is recorded in the 3d frame per second. The resolution of the recorded image is 1536 × 768. The total length of recorded video is around 80 minute (Fig. \ref{fig:Data}(a)).

\subsection{Observation model: discriminator to calculate o(t)}
Interaction motion is defined as $I^r_(\tau)$ containing L coordinates vectors from time point $\tau - L$ to $\tau$. The discriminator is implemented by convolutional neural network (CNN) trained by semi-supervise learning method, where fake interaction data $I^f$ is generated from real interaction data $I^r$ by negative sampling. Then the real interaction and fake interaction data is mixed for training. We used the network structure of CNN, which is Fully-connected layer (On 2nd axis 56 to 48), conv1(channel=16, kernel=[3,3], stride = 1, padding = [0,0]), conv2(channel=32, kernel=[3,3], stride = 2, padding = [1,0]), conv3(channel=64, kernel=[3,3], stride = 2, padding = [1,0]), Fully-connected layer (3200 to 128, dropout =0.5), and FC(128 to 1).

Before training the network, We normalized the mixed dataset to between 0 and 1, then augment data by adding noise in the temporal axis and spatial axis. Fake interaction data $I^f$ can be generated from real interaction data $I^r$ by shifting the time of $I^r_e (\tau)$  or $I^r_p (\tau)$ randomly. The objective of training a Discriminator is to minimize binary cross entropy.

The objective of training a Discriminator is to minimize binary cross entropy.
\begin{equation}
    Loss = -[y log F_d(I)+(1-y) log(1- F_d(I))]
\end{equation}

where $F_d(I)$ is the prediction on interaction motion $I$, y is the ground truth of whether interaction motion $I$ is real data or not, y = 1 when $I$ is real data, else y = 0. Fig. \ref{fig:Learning curve of discriminator}(a) shows that the interaction motion discriminator can achieve the accuracy of 60.5\%, which is better than us since We can not tell most the fake data when we watch them manually.

\subsection{Probabilistic Roadmap for Motion Sampling}
\begin{figure}
    \centering
    \includegraphics[width=1\linewidth]{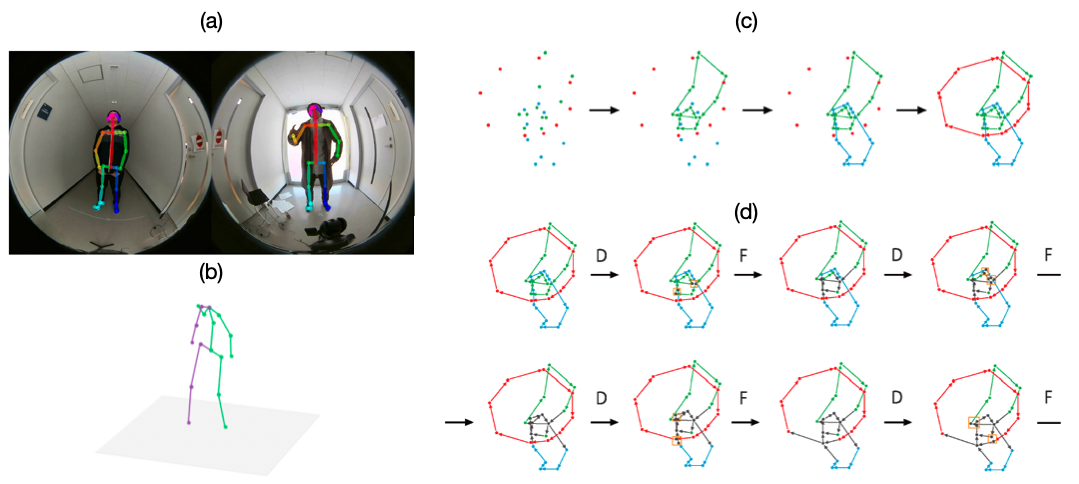}
    \caption{Data preparation. (a)Skeleton keypoints extracted by OpenPose. Participants (left side) and experi- menter (right side)(b)Result of 3d posture estimation(c)Process of initializing Roadmap(d)Process of fusing nodes in Roadmap}
    \label{fig:Data}
\end{figure}

In order to apply the sampled motion to robots and virtual agent, it is necessary to construct PRM by 3D skeleton keypoints. We use a 3D posture estimation method to estimate 3D skeleton keypoints for each 2D skeleton keypoints\cite{martinez2017simple}. The outcome is shown in Fig.\ref{fig:Data}(b). Roadmap is constructed by motion of experimenter-side in the dataset $C_e$. Since robots are typically controlled by joint angles, We construct dataset R by calculating the joint rotation for each frame in the 3D skeleton keypoints sequence. Coordinates vector $R_(\tau)$ stands for the joint angles of dataset R at time point $\tau$.

In learning phase, graph G = \{n, l\} is initialized by connecting nodes according to the consecutiveness of time point (e.g. from $R(\tau-1)$ to $R(\tau)$) as shown in Fig.\ref{fig:Data}(c). Probabilistic connections is established based on joint angular velocity $V(\tau)$ and joint angular acceleration $W(\tau)$. The joint angular velocity $V(\tau)$ and joint angular acceleration $W(\tau)$ is calculated by the difference of joint angles (e.g. $R(\tau-1) - R(\tau)$) and the difference of joint angular velocities (e.g. $V(\tau) -V(\tau-1)$)). Since the human motion must be fluent, the variation between two reachable nodes can be considered lower than some limitations.The relationship between joint angular velocity and joint angular acceleration is modeled by linear regression.



Openpose\cite{cao2017realtime,simon2017hand,wei2016convolutional}, is used to extract 2d skeleton keypoints sequences of human body from the videos. The output of OpenPose is quite noisy, so We remove outlier and resample data to 8 fps by linear interpolation. Since the human motion during interact can be considered always within some specific frequency, We use a low pass filter with cutoff frequency of 4 Hz to smooth data. 

The dataset C is consists of skeleton keypoints sequence of experimenters $C_p$ and participants $C_e$. $C_(\tau)$ stands for the coordinates vector at time point $\tau$, which contains the keypoints coordinates of participants $C_p(\tau)$ and experimenters $C_e(\tau)$. They are also shown in the left side and right side of Fig.\ref{fig:Data}(a) respectively.

\subsection{Training IO-HMM thorough EM algorithm}
\begin{figure}
    \centering
    \includegraphics[width=1\linewidth]{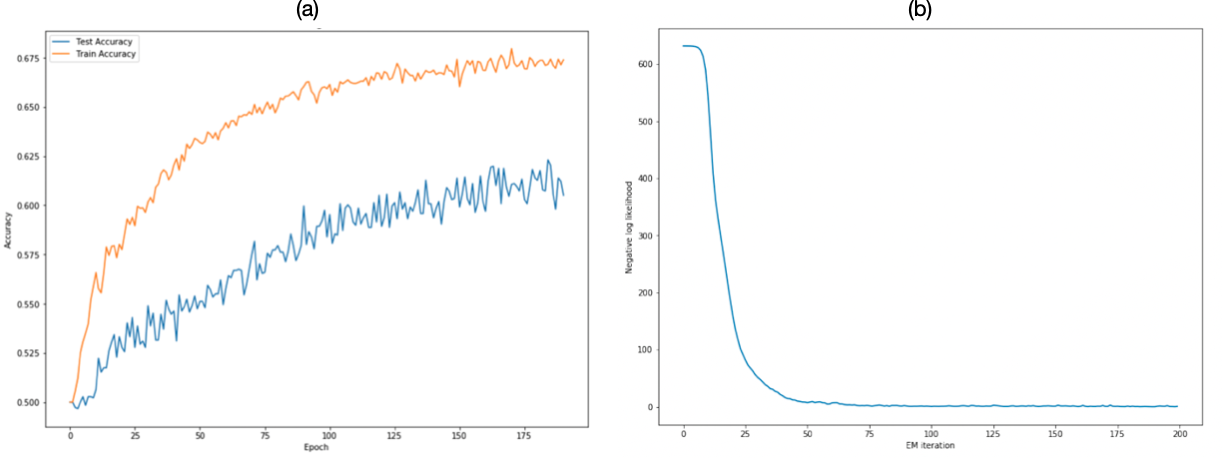}
    \caption{Learning curve (a)Learning curve of discriminator. Orange: Train accuracy, Blue: Test Accuracy (b)Learning curve of IO-HMM}
    \label{fig:Learning curve of discriminator}
\end{figure}

The input sequence A(T) and the output sequence O(T) is necessary to estimate parameters in Input-output Hidden Markov Model (IO-HMM) (Table.~\ref{table:data_type}). Data used in training IO-HMM is skeleton keypoints sequence of experimenter $C_e$ and the prediction of interaction motion discriminator $F_d(C)$ given dataset C as input sequence A(T) and output sequence O(T). The parameter $\theta_{EM}$ for emission model is a set of array where $\theta^i_{EM}$  denote the coefficients when the hidden state is $s_i$. The parameters $\theta$ of this probabilistic model is estimated to maximize likelihood L($\theta$,$O(t)$, $A(t)|C$) under the dataset C by Expectation-Maximization (EM) algorithm. Fig. \ref{fig:Learning curve of discriminator}(b) shows the learning curve of training IO-HMM and this figure suggests that this probabilistic model can fit our dataset.

\section{Experiments}
\subsection{Impression on the FEP based behavior}

\begin{figure}[htpb!]
    \centering
    \includegraphics[width=1\linewidth]{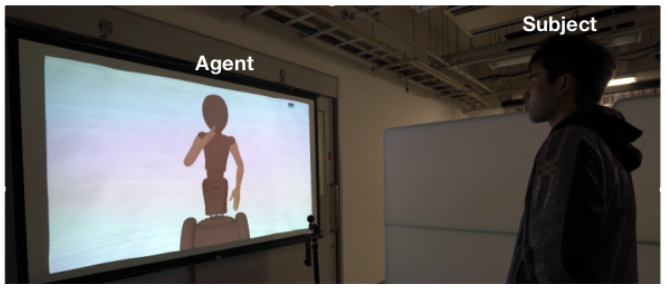}
    \caption{Evaluation experiment for three conditions}
    \label{fig:Experiment}
\end{figure}

To the purpose of evaluating impression on the Free Energy Principle (FEP) based behavior in human-perceptible level, We gather some experiment subjects and the impression is measured by the score they gave to the generated motion. The experiment subjects are indicated to bring an object back from the front of the projected human-size virtual agent and evaluate the impression on the agent. An omnidirectional camera is used for capturing the sensory input of the virtual agent. The snapshots of evaluation experiment for each condition is shown in Fig. \ref{fig:Experiment}. 

The impression is evaluated by three indexes, i) ``lifelikeness'', ii) ``human-likeness'' and iii) ``feeling of interacting with agent''. Three motion generation methods are compared in this experiment, i) ``Minimizing free energy'', ii) ``Sampling based on PRM'' and iii) ``Perlin Noise''. Experiment subject needs to answer three questions after each time they bring the object back. The order of motion generation method is determined randomly beforehand.

Motion generated by ``Minimizing free energy'' is implemented as motion generation system explained in section 3. At each time step, a set of possible motion is sampled from Probabilistic Roadmap (PRM) and the one with minimal free energy is selected. The motion generated by ``Sampling based on PRM'' is implemented as sampling motion from PRM randomly explained in chapter 3. The motion generated from ``Perlin Noise'' is implemented as adding noise to idle posture. In robotics, ``Perlin Noise'' has been used, added to joint angles, to increase the lifelikeness of robot movements and to generate idle motion. The experiment is conducted in a laboratory room of the university. The experiment subjects are bachelor students, master students and doctoral students of robotics major. We gathered 8 experiment subjects and repeat 5 times for each condition. 

\begin{figure*}
    \centering
    \includegraphics[width=1\linewidth]{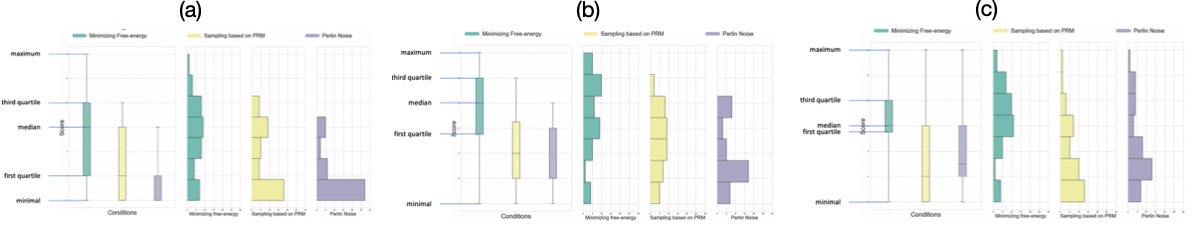}
    \caption{Impression on the FEP based behavior (a) Feeling of interaction. (b) Human-lileness (c) Life-likeness}
    \label{fig:Impression}
\end{figure*}

The subjective impression on three motion generation methods is shown in Fig.\ref{fig:Impression}. As explained in the last section, experiment subjects score the generated motion from 1 to 7. To express the difference between three motion generation method intuitively, the box plot on the left side of each figures shows the minimum value, first quartile, median value, third quartile value and maximum value of score for each motion generation method. On the right side of each figure, histograms describe how many people gave a specific score for motion generated by ``Minimizing free energy'', ``Sampling based on PRM'' and ``Perlin Noise''.

By using Mann-Whitney’s U test \cite{mann1947test}, The results suggest that  motion generated by ``Minimizing free energy'' shows a more ``feeling of interacting with agent'' than both ``Sampling based on PRM'' and ``Perlin Noise'' (p $<$ 0.01) (Fig.~\ref{fig:Impression}(a)). Moreover, data indicate that motion generated by ``Minimizing free energy'' shows more ``lifelikeness'' and ``human-likeness'' than both ``Sampling based on PRM'' and ``Perlin Noise'' (p $<$ 0.01) (Fig.~\ref{fig:Impression}(b)(c)). By watching the recorded video of evaluation experiment, it can be found that the intensity of motion may also affect the subjective impression. For instance, the intensity of motion generated by ``Perlin Noise'' and ``Sampling based on PRM'' is limited when comparing to ``Minimizing free energy''.

\subsection{Agents change humans behavior}


Then, to investigate whether or not the agent causes the pull-in, we gathered new subjects and experimented in the same environment as Fig.~\ref{fig:Experiment}. This time, we experimented with two conditions: one with a recording agent and the other with an agent interacting in real-time. For the recording condition, we used a video of the author interacting with the agent beforehand. Three subjects were used for each condition. Subjects entered a room where a human-size agent was projected and were asked to record the names of 10 fruits into a microphone in front of them. The subjects were instructed to start recording at their timing and to leave the room when they were finished and were not told that an agent was projected inside. We filmed the agent and the subject during their stay in the room using a video camera.

The time spent by the subject in the room where the agent was projected was measured. One of the subjects who interacted with the agent in real-time showed interest in the agent's gestures and was listening to the agent's gestures every time the name of fruit was mentioned during the task. After the task, the agent responded to the sound. After the task was over, the subject was observed vocalizing into the microphone and mimicking the agent's gestures to see if the agent responded to the sound.

\section{Discussion}
We implemented a free-energy based agent that makes action choices based on a set of self and opponent gestures. This agent improved the impression of interaction for the subject, as well as the impression of human-likeness and life-likeness. In addition, we found that subjects who confronted the agent paid more attention to the gestures generated by the agent and spent more time observing them, and some subjects were drawn into non-verbal interaction by imitating the gestures themselves.

In human-agent interaction, there have been many studies of agents that take only the other person's actions as input. The proposed model generates gestures by taking into account the set of actions of both the agent and the other person, so that the mutual attraction that occurs during interaction between humans can be realized in human-agent interaction. The agent's gestures were able to induce a change in the behavior of the person. It is expected to open the door to the exciting challenge of agents that actively work with people and encourage them to change their behavior.

\section{Acknowledgments}
This work was supported by JST Moonshot R\&D Grant Number JPMJMS2011.

\bibliography{AROB.bib} 
\bibliographystyle{plain} 
\end{document}